\documentclass[twocolumn]{revtex4}
\usepackage{}
\usepackage{epsfig,latexsym,amssymb}
\usepackage{latexsym}
\usepackage{amssymb}
\usepackage{amsfonts}
\usepackage{comment}
\usepackage{graphicx}
\usepackage{verbatim}

\newcommand{\be}{\begin{equation}}
\newcommand{\ee}{\end{equation}}
\newcommand{\bea}{\begin{eqnarray}}
\newcommand{\eea}{\end{eqnarray}}

\begin{document}

\title{Sub-MeV Bosonic Dark Matter, Misalignment Mechanism and Galactic Dark Matter Halo Luminosities}

\author{Qiaoli Yang$^{1}$\footnote{qiaoli\_yang@hotmail.com} and Haoran Di$^{2}$\footnote{haoran\_di@yahoo.com}}
\affiliation{$^{1}$Department of Physics, Jinan University, Guangzhou 510632, China\\$^{2}$School of Physics, Huazhong University of Science and Technology, Wuhan 430074, China}

\begin{abstract}
We explore a scenario that dark matter is a boson condensate created by the misalignment mechanism, in which a spin 0 boson (an axion-like particle) and a spin 1 boson (the dark photon) are considered, respectively. We find that although the sub-MeV dark matter boson is extremely stable, the huge number of dark matter particles in a galaxy halo makes the decaying signal detectable. A galaxy halo is a large structure bounded by gravity with a typical $\sim10^{12}$ solar mass, and the majority of its components are made of dark matter. For the axion-like particle case, it decays via $\phi\to \gamma\gamma$, therefore the photon spectrum is monochromatic. For the dark photon case, it is a three body decay $A'\to\gamma\gamma\gamma$. However, we find that the photon spectrum is heavily peaked at $M/2$ and thus can facilitate observation. We also suggest a physical explanation for the three body decay spectrum by comparing the physics in the decay of orthopositronium. In addition, for both cases, the decaying photon flux can be measured for some regions of parameter space using current technologies.
\end{abstract}

\maketitle

{\itshape Introduction:}
Mounting evidence shows that dark matter constitutes about one third of the total energy density or more than 80\% of the matter density of our universe. In galaxy halos, the mass fractions of dark matter are even more dominated which can be as high as 95\% \cite{Battaglia:2005rj,Springel:2005nw,Smith:2006ym,Duffy:2008pz,Okabe:2009pf,Guo:2009fn,Postman:2011hg}. Although the existence of dark matter is widely accepted, the exactly nature of dark matter particles are still little known \cite{Queiroz:2016sxf} besides that they are non-baryonic, weakly interacting and stable. There are many possible dark matter particle candidates, and two of them, i.e., the axion-like particles $\phi$ \cite{Preskill:1982cy,Abbott:1982af,Dine:1982ah,Berezhiani:1989fu,Ipser:1983mw,Svrcek:2006yi,Simet:2007sa,DeAngelis:2007dqd,Arvanitaki:2009fg,Jaeckel:2010ni,DeAngelis:2011id,Ringwald:2012hr,Marsh:2015xka} and the hidden massive U(1) vector boson $A_{\mu}'$ \cite{Ahlers:2007rd,Redondo:2008ec,Jaeckel:2008fi,Abel:2008ai,Goodsell:2009xc,Mirizzi:2009iz,Bullimore:2010aj,An:2013yfc,Redondo:2013lna}, are especially interesting because both can be produced from the misalignment mechanism \cite{Preskill:1982cy,Abbott:1982af,Dine:1982ah,Ipser:1983mw,Svrcek:2006yi,Nelson:2011sf,Arias:2012az}. The misalignment mechanism does not require any particular interactions between the dark matter particles and the Standard Model sector as far as dark matter particles are bosons and massive in the very early universe. Therefore, the spin 0 case can be the axion or axion-like particle, in which case one of the "string axions" accounts for the majority of the dark matter, and the spin 1 case can be the dark photon.

If the dark matter is composed of axion-like particles or dark photons, then the dark matter could be a boson condensate \cite{Sikivie:2009qn} with a high number density. As these cosmic dark matter particles are not completely stable, they can decay into three (from the dark photon) or two (from the axion-like particle) photons. Certainly, to be a proper dark matter candidate the lifetime of the particle should be beyond the age of the universe, however, the luminosity of decaying photons from a dark matter halo can be measured due to the huge number of particles in a halo. In addition, we find that the signal of the decayed photons is very unique. For the axion-like particle case, it is a two body decay $\phi\to 2\gamma$, so the energy spectrum of the photons is monochromatic at $E=M_{\phi}/2$, where $M_{\phi}$ is the mass of the axion-like particle. For the dark photon case, although it is a three body decay, the energy spectrum is heavily peaked (see Fig.2) according to our calculation. The physical reason for this three body decay spectrum is understandable if we compare it with the spectrum of the orthopositronium decay \cite{Manohar:2003xv}, which we discuss in the following section.

{\itshape Dark Matter Models and Misalignment Mechanism:}
Hypothetical particles in the hidden sector are generally interacting very weakly with the standard model sector. The hidden U(1) vector boson was originally proposed to be a massless particle that carried a long range force on dark matter. It was later realized that the massive vector bosons themselves could be a very good candidate for dark matter. Here, let us use $A'_{\mu}$ to denote the hidden sector U(1) vector field. The Lagrangian is:
\bea
{\cal L}&=&-{1 \over 4}(F^{\mu\nu}F_{\mu\nu}+F'^{\mu\nu}F'_{\mu\nu})-{M_{A'}^2\over 2} A'_{\mu}A'^{\mu}\nonumber\\
&-&C_V\bar f \gamma^\mu f A'_{\mu}+...  ~~,
\eea
where $M_{A'}$ is the mass of the dark photon, $F_{\mu\nu}$ and $F'_{\mu\nu}$ represent the field strength tensor of $A_{\mu}$ and $A'_{\mu}$ respectively. $C_V$ is the vector couplings of the dark photon due to kinetic mixing with ordinary photons and $f$ is the Standard Model fermions. In this paper, we assume that the dark photon mass term $M_{A'}$ is a Stueckelberg mass \cite{Feldman:2007wj} instead of a Higgs mass. The Stueckelberg mass can be naturally induced from string compactifications, thus the dark photons can be safely regarded as massive without passing through a spontaneous symmetry breaking (SSB) phase transition until the very early universe. The misalignment mechanism then applies in the scenario. Interestingly, nature seems prefer to use every renormalizable Lagrangian term, yet the Stueckelberg mass is absent in the Standard Model. If the Stueckelberg mass exists, it applies only to abelian gauge fields as far as renormalizability is required, thus we may suspect that dark photons do exist in our universe.

The initial value of the $A'_{\mu}$ field is assumed to be some random nonzero value, and the inflation creates a spatially uniform field, thus $\partial_i A'_{\mu}\sim 0$. In a Friedmann-Robertson-Walker (FRW) universe and the cosmic frame, the equation of motion of the vector field \cite{Nelson:2011sf,Arias:2012az} implies that $A'_0=0$ if $M_{A'}\neq 0$, and
\bea
\partial_t^2 \vec A'&+&3H(t) \partial_t\vec A'\nonumber\\
+(M_{A'}^2&+&\dot H(t)+2H(t)^2)\vec A'=0,
\label{eq2}
\eea
where $H(t)$ is the Hubble parameter. After the universe enters a radiation dominated era Eq.[\ref{eq2}] reduces to $\partial_t^2 \vec A'+(3/2t) \partial_t \vec A'+M_{A'}^2\vec A'=0$. The solution of the equation is $A'_i=C_1J_{1/4}(M_{A'}t)/t^{1/4}+C_2Y_{1/4}(M_{A'}t)/t^{1/4}$, where $J_N$, $Y_N$ are the Bessel functions and $C_1$, $C_2$ depend on the initial conditions. The solution can be explained intuitively. When $H(t)\gg M_{A'}$, the field strength is decreasing without any oscillations. Thus no particles are observed during this period, only a uniform field in the universe. After the Hubble constant decreases to a point that $3H(t)\sim 2M_{A'}$, the field begins to oscillate. The amplitude of the initial oscillation determines the energy density of the new born particles so in this scenario we have a non-thermal creation mechanism. After $H\ll M_{A'}$ the energy density of the vector field $\rho\sim A'^2$ decreases as $1/a^3(t)$ therefore behaves like dust. With a proper initial value and particle mass combination, the mechanism can provide the right energy density of the dark matter. As the transition happens at the same time in the universe, the particles are in a coherent state with small fluctuations and are very cold due to their coherent nature regardless of their mass.

The axion is introduced to provide a natural solution to the strong CP problem \cite{Peccei:1977hh,Weinberg:1977ma,Wilczek:1977pj,Kim:1979ifz,Shifman:1979if,Dine:1981rt}. Generally speaking, to solve the strong CP problem, one may introduce a new $U(1)$ symmetry called the Peccei-Quinn symmetry. The Peccei-Quinn symmetry is spontaneously broken when the energy is below the scale $\Lambda\sim f_a\sim 10^{12}{\rm GeV}$ and subsequently is explicitly broken due to non-perturbative quantum chromodynamics (QCD) effects after the energy drops below the QCD scale. The SSB gives a goldstone boson and the explicitly symmetry breaking gives the goldstone boson a small non-zero mass. This pseudo-goldstone boson is now known as the axion. The axion-like particles \cite{Becker:1995kb,Kallosh:1995hi,Douglas:2006es,Svrcek:2006yi,Arvanitaki:2009fg} are zero Kaluza-Klein modes created from compactifying antisymmetric tensor fields on closed cycles of space-time manifold. These zero modes acquire a potential from non-perturbative effects on the cycle and therefore obtain a small mass. Although the compactifications give rise to many axion-like particles, we consider a scenario in which one of them composes the majority of dark matter. Both axions and axion-like particles have similar properties in low energy four-dimensional effective field theory, except in the case of axions the mass and the SSB scale are closely related. We can write the general Lagrangian for axions and axion-like particles as follows:
\bea
{\cal L}&=&-{1 \over 4}F^{\mu\nu}F_{\mu\nu}-{1\over 2}\partial_{\mu}\partial^{\mu}\phi-{M_{\phi}^2\over 2}\phi^2+{\alpha g\over2\pi \Lambda}\phi F_{\mu\nu}F_{\alpha\beta}\epsilon^{\mu\nu\alpha\beta}\nonumber\\
\eea
where $\Lambda$ is the order of symmetry breaking energy scale, $g$ is a model dependent factor of order one which we will neglect in the following discussion, $\alpha$ is the fine structure constant, and $\phi$ is the particle field. For the QCD axions we have \cite{Sikivie:2006ni}:
\be
M_{\phi}\sim m_{\pi^0}f_{\pi}/\Lambda\simeq 6.00*10^{-6}{\rm eV}({10^{12}{\rm GeV}/\Lambda})~~.
\label{mass}
\ee
In addition, inflation can produce isocurvature perturbations of scalar fields which are highly constrained from the observations of cosmic microwave background radiations. As such we have additional limitations for the axion \cite{Turner:1985si,Lyth:1989pb,Turner:1990uz,Lyth:1992tx,Burns:1997ue,Fox:2004kb,Beltran:2006sq,Hertzberg:2008wr,Estevez:2016keg}.

Cold axion-like particles can be created from the misalignment mechanism \cite{Preskill:1982cy,Abbott:1982af,Dine:1982ah,Ipser:1983mw,Svrcek:2006yi}. The equation of motion of $\phi$ in a FRW universe is $\partial_t^2 \phi+3H \partial_t \phi+M_{\phi}^2\phi=0$, so the production mechanism is similar to that of the dark photon. In addition, the initial field value of axion-like particles is constrained as $\phi/\Lambda$ is corresponding to a phase value order of one. Let us use $t_1\sim 1/M_{\phi}$ to denote the time that axion-like particle field starts to oscillate. The energy density of axion-like particles to the critical energy density during the matter dominated era is thus:
\bea
{\rho_a\over \rho_c}\sim {1\over2}M_{\phi}^2\Lambda^2\left ({a(t_1)\over a(t)}\right )^3/\rho_c\approx 3\pi G\sqrt M_{\phi}\Lambda^2\sqrt t_e<1,
\label{density}
\eea
where $a$ is the FRW scale factor and $t_e$ is the matter-radiation equality time. so we have a lower boundary in Fig.4.

As the Compton wavelength of dark matter should be no longer than 1kpc to allow for the structure formation on the kpc scale \cite{Nelson:2011sf}, both dark photons and axion-like particles have a weak bound on the mass, which is $M>10^{-24}$eV. In addition, the survival of dark matter requires particles to have lifetimes longer than the age of the universe. These constraints are shown in Fig.3 and Fig.4.

{\itshape Decaying Signals from Dark Matter Halos:}
Sub-MeV massive dark photons are not completely stable because they can decay via channel $\gamma'\to3\gamma$ due to the vector interaction. The axion-like particles can decay via $\phi\to 2\gamma$. The decay rate for the dark photon is:
\bea
\Gamma_{A'}&=& 2.725\times 10^{-15}{C_V^2M_{A'}^9\over m_e^8}\nonumber\\
&=&0.895*10^{-45}{\rm seconds}^{-1}( {M_{A'}\over {\rm eV}})^9C_V^2~~,
\eea
where $m_e$ is the mass of electrons. The decay rate of the axion-like particle $\Gamma_{\phi}$ is:
\bea
\Gamma_{\phi}&=&{\alpha^2\over4\pi^3\Lambda^2 }M_{\phi}^3\nonumber\\
&=&6.53*10^{-10}{\rm s^{-1}}({M_{\phi}\over{\rm eV}})^3({{\rm GeV}\over \Lambda})^2~~.
\eea
\begin{figure}
\begin{center}
\includegraphics[width=0.4\textwidth]{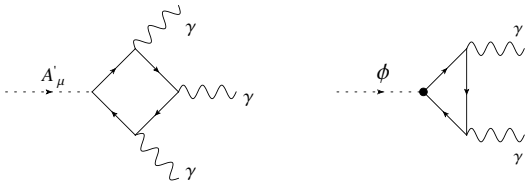}
\caption{The decay of a dark photon $A_{\mu}'$ and of an axoin-like particle $\phi$.}
\end{center}
\end{figure}

The axion-like particle decay is a two body decay and thus the resultant photons are monochromatic in the energy spectrum, with a frequency at $\omega=M/2$. The dark photon decay is a three body decay, and the spectrum of the decaying photons is calculated and is shown in Fig.2.
\begin{figure}
\begin{center}
\includegraphics[width=0.4\textwidth]{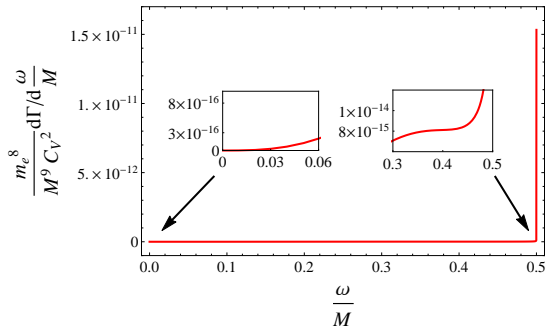}
\caption{The decay spectrum of the dark photon.}
\end{center}
\end{figure}
An interesting feature of the dark photon decay is that although it is a three body decay, which typically results to a smooth spectrum, e.g. in the $\beta$ decay, the decaying spectrum of the dark photon is however heavily peaked at $\omega=M/2$. We find that $36.1\%$ decaying energy lies in a frequency range $[0.9*M/2,~ M/2]$.

The physical reason for the dark photon decay spectrum sharply peaking at $M/2$ is due to the properties of the vector bosons and can be understood when we compare it with the orthopositronium decay spectrum. In the low energy end, both spectrums vanish linearly due to Low's theorem on soft photon emissions. In the $E\to M/2$ end, both spectrums rise sharply, as the phase space for a two photon decay is far more allowed than that for a three photon decay. Therefore, in the three photon decay, one infrared photon plus two $M/2$ photons is the most favorable situation. For the orthopositronium case, the phase space is much more tolerated as the initial state is different from a point particle. Therefore the spectrum of orthopositronium decay is much less peaked.

\begin{figure}
\begin{center}
\includegraphics[width=0.4\textwidth]{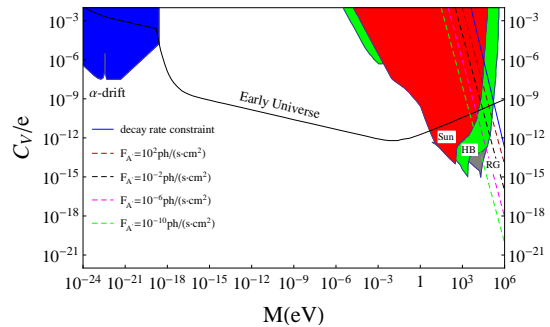}
\caption{The constraints and the equal flux lines of the sub-MeV dark photon. The blue area is excluded from the drift of the fine structure constant and the early universe behavior imposes an upper limit for the coupling $C_V$ \cite{Nelson:2011sf}. The red, green and gray areas are ruled out by the lifetime of stars (the sun, horizontal branch stars, red giants) \cite{Goodsell:2009xc,Bloch:2016sjj}. The lifetime of the dark photons impose a decay rate constraint for the upper bound of the mass. The structure formation imposes a lower bound of the mass $M_{A'}>10^{-24}{\rm eV}$ \cite{Nelson:2011sf}. The dashed lines are the equal light flux lines, showing that the photons are emitted from the M31 dark matter halo and received on earth.}
\end{center}
\end{figure}
\begin{figure}
\begin{center}
\includegraphics[width=0.4\textwidth]{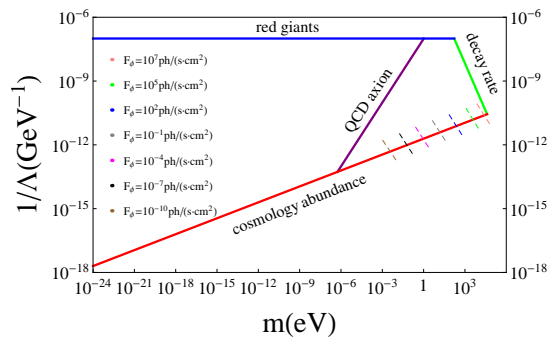}
\caption{The constraints and the equal flux lines of axion-like particles. The upper area is excluded from the stellar evolution constraint \cite{Ayala:2014pea,Giannotti:2015kwo}. The decay rate of the axion-like particle imposes an upper limit of the mass. The structure formation imposes a lower bound of the mass $M_{\phi}>10^{-24}{\rm eV}$. The lower area is excluded due to Eq.(\ref{density}), as it would produce too much dark matter. For the QCD axion, the mass and the decay constant is related by Eq.(\ref{mass}). The dashed lines are the equal light flux lines showing that the photons are emitted from the M31 dark matter halo and received on earth. Note that for an axion-like particle to be a substantial fraction of dark matter, its parameters should lie around the cosmology abundance line.}
\end{center}
\end{figure}

The spectrum of the decaying photons observed on Earth depends on several astronomical factors and there are four major sources that distort the spectrum: The first is the velocity dispersions $\delta v$ of the dark matter particles in a halo which lead to a $(1+\delta v/C)/(1-\delta v/C)\sim 2\delta v/C$ spectrum spread. The second is the gravitational red shifting due to local density profiles. The third is the kinetic movement of dark matter halos relative to the observer, and the fourth is the cosmological red shift due to expanse of the universe. Assuming the dark matter particles remain non-relativistic, $\delta v\ll 1$, the spectrum spread is small. Furthermore, the gravity induced spectrum spread should be small assuming a standard dark matter halo density profile. In addition, if the total mass of a source halo is similar to our galaxy halo, then the center frequency red shift due to photons moving in and out of halo gravity wells should be largely canceled. Finally, let us use $z_k$ and $z_c$ to denote the magnitude of the red shift due to movement of the halos and due to the cosmological red shift respectively. As $z_k$ and $z_c$ are small for nearby halos, the observed spectrum is peaked at $(1+z_k+z_c)^{-1}M/2$.

To be a proper candidate for dark matter, the lifetime of particles $\tau=1/\Gamma$ should be larger than the age of the universe. However, due to a huge particle number $N$ in a dark matter halo, the luminosity $L$ from decaying dark photons or decaying axion-like particles of a dark matter halo may be detectable. A galaxy halo is a heavy gravity bound structure with a typical mass ranging from $10^{11}$ to $10^{13}$ solar masses, and the most \cite{Battaglia:2005rj,Springel:2005nw,Smith:2006ym,Duffy:2008pz,Okabe:2009pf,Guo:2009fn,Postman:2011hg} of the halo mass is composed of dark matter. Let us consider a dark matter halo with a mass $M_h$. The luminosity of the dark photon halo is:
\bea
L_{A'}&=&{N_{A'}\times M\over \tau}=0.95{M_h\over \tau}\nonumber\\
&=&3.96*10^{-25}{M_h\over M_{\odot}}({M\over {\rm eV}})^9C_V^2L_{\odot}~~.
\eea
For the axion-like particle dark matter halo the luminosity is:
\bea
L_{\phi}&=&{N_{\phi}\times M_{\phi}\over \tau}\nonumber\\
&=&28.9*10^{10}{M_h\over M_{\odot}}({M_{\phi}\over {\rm eV}})^3({{\rm GeV}\over \Lambda})^2L_{\odot}~~,
\eea
where $M_\odot$ and $L_\odot$ denote the solar mass and the solar luminosity respectively. For QCD axions, the mass and the energy breaking scale are closely related by Eq.(\ref{mass}). As such we have $L_{\phi}\propto(M_h/M_{\odot})*(M_\phi/{\rm eV})^5L_{\odot}$ which is consistent with the literature \cite{Kephart:1986vc,Bershady:1990sw,Grin:2006aw}.

The observed light flux at a distance is:
\be
F={L\over 4\pi d_L^2}
\ee
where $d_L$ is the luminosity distance. For nearby galaxy halos it is sufficient to use $d_L\approx z_cc/H_0$, where $c$ is the speed of light. The Hubble constant $H_0\approx 73{\rm (km/s)/Mpc}$ \cite{Riess:2016jrr}, so we have $d_L\approx 4.11*10^3z_c$Mpc. The observed flux within a frequency range $[0.9M/2,~M/2]$ from a dark photon galaxy halo with a cosmological red-shift $z_c$
is then:
\bea
F_{A'}&=&2.69*10^{-56}{M_h\over M_{\odot}}({M\over {\rm eV}})^9C_V^2z_c^{-2}{\rm (W/cm^2)}\nonumber\\
&=&3.36*10^{-37}{M_h\over M_{\odot}}({M\over {\rm eV}})^8C_V^2z_c^{-2}{\rm ph/(s~cm^2)}.
\eea
For the axion-like particle dark matter halo the observed flux is:
\bea
F_{\phi}&=&5.46*10^{-20}{M_h\over M_{\odot}}({M_{\phi}\over {\rm eV}})^3({{\rm GeV}\over \Lambda})^2z_c^{-2}{\rm (W/cm^2)}\nonumber\\
&=&0.683*{M_h\over M_{\odot}}({M_{\phi}\over {\rm eV}})^2({{\rm GeV}\over \Lambda})^2z_c^{-2}{\rm ph/(s~cm^2)}~.
\eea

The NGC 224 or Messier 31 galaxy is a major neighbor galaxy near the earth. The distance from NGC224 to the earth is 0.778Mpc \cite{Ribas:2005uw} which corresponds to a $z_c=1.89*10^{-4}$. The mass of NGC 224 is about $1.5*10^{12}$ \cite{Penarrubia:2014oda} solar masses. Star formations in the NGC 224 disk are near inactive, and the galactic neutral hydrogen, molecular hydrogen and dust combined counts about $10^{10}$ solar masses or below 1\% of the total masses \cite{Draine:2013mfa}. Therefore the spectrum distortions should be small. We expect the received signal to be a narrow line that is peaked at $(1+z_k+z_c)^{-1}M/2$ where $(z_k+z_c)\sim -10^{-3}$ for the NGC 224. If the dark matter is composed by the dark photon, then the luminosity of NGC 224 is $5.98*10^6L_{\odot}({M\over60{\rm keV}})^9({C_V\over10^{-12}})^2$ and the flux within $[0.9M/2,~M/2]$ received on earth is $2.37*10^{-3}({M\over 60{\rm keV}})^8({C_V\over10^{-12}})^2{\rm ph/(s~cm^2)}$. If the dark matter is composed of the axion/axion-like particles, then the luminosity and the flux are $0.434L_{\odot}({M\over{\rm eV}})^3(10^{12}{{\rm GeV}\over\Lambda})^2$ and $2.87*10^{-5}({M\over{\rm eV}})^2(10^{12}{{\rm GeV}\over\Lambda})^2{\rm ph/(s~cm^2)}$ respectively.

{\itshape Conclusions:}
Boson condensate produced from the misalignment mechanism is a viable candidate for dark matter. The spin 0 axions/axion-like particles and the spin 1 dark photons are especially well motivated. The huge particle number in a galactic halo makes the decaying signal detectable for some regions of parameter space. Both types of dark matter candidate have a very sharp spectrum line peaked at $(1+z_c+z_k)^{-1}M/2$. A scan of the relative spectrum of nearby galactic halos combined with complementary laboratory experiments may reveal the nature of dark matter.

{\itshape Acknowledgments:} Q. Yang would like to thank XiangSong Chen, JianWei Cui, WeiTian Deng, Bo Feng, YunGui Gong and YunQi Liu  for their helpful discussions. This work is partially supported by the Natural Science Foundation of China under grant No. 11305066.

\end{document}